\documentstyle[aps]{revtex}
\begin{document}
%\begin{titlepage}
\draft
\title{Self-gravitating Line Sources of Weak Hypercharge}
\author{ T. Dereli}
%Department of Physics, Middle East Technical University\\
%06531 Ankara, Turkey\\ 
%{\footnotesize E.mail: tekin@dereli.physics.metu.edu.tr}\\ \\
\address{Department of Physics, Middle East Technical University\\
06531 Ankara, Turkey\\ 
{\footnotesize E.mail: tekin@dereli.physics.metu.edu.tr}}
%\author{ Robin W Tucker}
%Department of Physics, University of Lancaster\\
%Bailrigg, Lancs. LA1 4YB, UK\\ 
%{\footnotesize E.mail: r.tucker@lancaster.ac.uk}}
\author{Robin W Tucker}
\address{Department of Physics, University of Lancaster\\
Bailrigg, Lancs. LA1 4YB, UK\\ 
{\footnotesize E.mail: r.tucker@lancaster.ac.uk}}
\date{15 August 1998}
\maketitle
\begin{abstract}
We explore the role of the Cremmer-Scherk mechanism in the context
of low energy effective string theory by coupling the antisymmetric
3-form gauge field to an Abelian gauge potential carrying
weak hypercharge. The theory admits a class of exact 
self-gravitating solutions in the spontaneously broken phase
in which dual fields acquire massive perturbative modes. 
Despite the massive nature of these fields they admit non-perturbative
progressive longitudinal modes that together with
pp-type gravitational waves travel in a direction of a line source at
the speed of light.
\end{abstract}
%\end{titlepage}
%\newpage
\vskip 5mm

Considerable effort has been devoted to the search for classical string-like
solutions in relativistic field theories. Such solutions range from 
the pioneering work on vortices as models for dual strings [1]
to more recent investigations on the properties of global
and superconducting cosmic strings [2]-[4].
A common feature in recent work has been the role played by singular sources
as models for the strings themselves.
Such source descriptions often lend themselves to a formulation
in terms of de Rham
periods. Thus in the Higgs vacuum of the global Abelian
Higgs model [3], the phase $\theta$ of a complex scalar field
satisfies the massless field equations $d*d\theta = 0$ in a regular
space-time domain. In such a domain one may introduce a 2-form
potential $\hat{B}$ by $d\hat{B} = *d\theta$. Classical sources enter the
theory as solutions with $\int_C d\theta = 2\pi$ for some closed space-like
curve $C = \partial \Sigma_2$ bounding a space-like disc $\Sigma_2$.
For such solutions, $\hat{B}$ can be promoted to a distribution on space-time
satisfying $d*d\hat{B} = 2\pi \delta$ with $\int_{\Sigma_2} \delta = 1$.
One then identifies the solution as a line source threading $C$
at each instant and such "axionic" strings have interesting
cosmological implications [5].

In a recent note we have suggested that fields arising in low energy
effective string actions may have consequences for the standard
model of the electroweak interactions [7]. By coupling the antisymmetric
3-form gauge field $H$ to an Abelian gauge potential 1-form $A$  
carrying weak hypercharge via a gauge covariant derivative
of the standard Higgs weak isospinor, we showed explicitly
how the masses of the $W^{\pm}, Z^0$ could depend on this coupling.
A salient feature of this generalised
Cremmer-Scherk mechanism [6] was the manner in which the $H$ field
became assimilated   into the physical degrees of freedom
of the vector bosons via a spontaneous breakdown
of a local gauge symmetry. Since the $H$ field
along with the dilaton $\phi$ is thought to have 
implications for cosmology, it is of interest to explore the
gravitational sector of the low energy effective string
action in the presence of the Cremmer-Scherk interaction.
Although the model discussed in Ref.[7] involves the full
non-Abelian $SU(2) \times U(1)$ gauge theory of the electroweak
standard model, we shall here restrict attention to a single
local Abelian  internal symmetry gauge group
for simplicity but retain the hypercharge interpretation.
The Cremmer-Scherk mechanism is controlled by a coupling
constant $\lambda$   and we are interested
in the phase with $\lambda \neq 0$.
Thus to lowest order in string fields we investigate
the dynamics derived from the action density 4-form
%\begin{eqnarray}
%\Lambda[{\bf g}, \phi, A, B] &=& \kappa {\cal R}*1 - \frac{(2\alpha-3)}{4}
%d\phi \wedge *d\phi \nonumber \\
%& & + \frac{1}{2} e^{-2\phi} dB \wedge *dB
%+ \frac{1}{2} e^{-2\phi} dA \wedge *dA
%+ \lambda A \wedge dB
%\end{eqnarray}
\begin{equation}
\Lambda[{\bf g}, \phi, A, B] = \kappa {\cal R}*1 - \frac{(2\alpha-3)}{4}
d\phi \wedge *d\phi \nonumber \\
 + \frac{1}{2} e^{-2\phi} dB \wedge *dB
+ \frac{1}{2} e^{-2\phi} dA \wedge *dA
+ \lambda A \wedge dB
\end{equation}
%when $\lambda = 0$,
where $A$ is a 
1-form, $B$ a 2-form, $\phi$ the dilaton 0-form
on spacetime $M$ with a metric $
{\bf g} = \eta_{ab} e^{a} \otimes e^{b}
$, curvature scalar $\cal R$
 and associated Hodge map $*$. 
The field equations derived from (1) by varying
$A, B, \phi, {\bf g}$, respectively, are
\begin{equation}
d({e^{-2\phi}} *dA) + \lambda dB = 0,
\end{equation}
\begin{equation}
d({e^{-2\phi}} * dB) - \lambda dA = 0,
\end{equation}
\begin{equation}
d*d\phi = \frac{2}{(2\alpha-3)} e^{-2\phi}(dB \wedge *dB
 + dA \wedge *dA),
\end{equation}
\begin{equation}
\frac{\kappa}{2} R_{bc} \wedge *(e_{a} \wedge e^{b} \wedge e^{c}) =
\tau_{a}[\phi] + \tau_{a}[A] + \tau_{a}[B],
\end{equation}
where
\begin{eqnarray}
\tau_{a}[\phi] &=& \frac{(2\alpha-3)}{4} (\iota_{a}d\phi *d\phi
+ d\phi \iota_{a}*d\phi) \nonumber \\
\tau_{a}[A] &=& \frac{1}{2}e^{-2\phi}(\iota_{a}dA \wedge *dA - dA \wedge
\iota_{a}*dA) \nonumber \\
\tau_{a}[B] &=& \frac{1}{2} e^{-2\phi}(\iota_{a}dB \wedge *dB
+ dB \wedge \iota_{a} *dB)
\end{eqnarray}
in terms of the interior operator with $\iota_{a}(e^b) = \delta_{a}^{b}$.
In a regular source-free domain of space-time
(2) and (3) imply 
\begin{equation}
d\tilde{A}= \lambda e^{2\phi} *\tilde{B}, 
\end{equation}
\begin{equation}
d\tilde{B}= \lambda e^{2\phi} *\tilde{A}, 
\end{equation}
in terms of the variables $\tilde{A} = A - \frac{1}{\lambda}df_0$,
$\tilde{B} = B - \frac{1}{\lambda}df_1$ in the gauge equivalence classes
$[A]$ and $[B]$, respectively.
One may  fix  gauges by taking solutions with particular $f_0$ and $f_1$.
Using (7) or (8) the entire theory can be recast in terms of either
the fields $\{{\bf g}, \phi, \tilde{A}\}$ or the fields
$\{{\bf g}, \phi, \tilde{B}\}$, and the two descriptions refer
to dual sectors of the same theory.
Moreover, in terms of the $\{{\bf g}, \phi, \tilde{A}\}$ description
the theory admits vector fields satisfying a generalised
Einstein-dilaton-Proca system:
\begin{equation}
d(e^{-2\phi} *d\tilde{A}) + {\lambda}^{2} e^{2\phi} *\tilde{A} = 0,
\end{equation}
\begin{equation}
d*d\phi = - \frac{2\lambda^2}{(2\alpha-3)} e^{2\phi}\tilde{A} \wedge *\tilde{A}
+ \frac{2}{(2\alpha-3)} e^{-2\phi} d\tilde{A} \wedge *d\tilde{A},
\end{equation}
\begin{equation}
\frac{\kappa}{2} R_{bc} \wedge *(e_{a} \wedge e^{b} \wedge e^{c}) =
\tau_{a}[\phi] + \tau_{a}[\tilde{A}] + \frac{\lambda^2}{2} e^{2\phi}
(\iota_{a}*\tilde{A} \wedge \tilde{A} + 
*\tilde{A} \wedge \iota_{a}\tilde{A}).
\end{equation}
It is clear how in the absence of gravitation the vector field $\tilde{A}$
acquires massive propagating modes. 
In a similar manner the theory admits a description in terms of
$\{{\bf g}, \phi, \tilde{B}\}$ satisfying the generalised
Einstein-dilaton-massive-Kalb-Ramond system:
\begin{equation}
%edit
d(e^{-2\phi} *d\tilde{B}) - {\lambda}^{2} e^{2\phi} *\tilde{B} = 0,
\end{equation}
\begin{equation}
%edit
d*d\phi = - \frac{2\lambda^2}{(2\alpha-3)} e^{2\phi}\tilde{B} \wedge *\tilde{B}
+ \frac{2}{(2\alpha-3)} e^{-2\phi} d\tilde{B} \wedge *d\tilde{B}.
\end{equation}
\begin{equation}
\frac{\kappa}{2} R_{bc} \wedge *(e_{a} \wedge e^{b} \wedge e^{c}) =
\tau_{a}[\phi] + \tau_{a}[\tilde{B}] + \frac{\lambda^2}{2} e^{2\phi}
(*\tilde{B} \wedge \iota_{a}\tilde{B} -
\iota_{a}*\tilde{B} \wedge \tilde{B}).
\end{equation}
 
Working with the fields $\{{\bf g}, \phi, \tilde{A}\}$, 
and restricting to cylindrical symmetry
we seek solutions for $\tilde{A}$ and $\phi$ with the metric
\begin{equation}
{\bf g} = du \otimes dv + dv \otimes du +
d\rho \otimes d\rho + \rho^{2} d\psi \otimes d\psi
+ 2 {\cal{H}}(u,\rho) du \otimes du,
\end{equation}
in a coordinate system $(u, v,  \rho, \psi)$. 
 We take $\cal{H}$ to have the form
\begin{equation}
{\cal{H}} = {f(u)}^{2} h(\rho)
\end{equation}
and
\begin{equation}
\tilde{A} = f(u) \beta(\rho) du
\end{equation}
with the dilaton constant,
\begin{equation}
\phi = \phi_0.
\end{equation}
It follows from (7) that the corresponding solution for $\tilde{B}$
will have the form
\begin{equation}
\tilde{B} = -\frac{e^{-2\phi_0}}{\lambda} f(u) {\beta}^{\prime}(\rho)
\, du \wedge \rho d\psi .
\end{equation}
The equations (9), (10) and (11) are satisfied provided 
the functions $\beta(\rho)$ and $h(\rho)$ solve
\begin{equation}
\beta^{\prime \prime} +\frac{1}{\rho} \beta^{\prime} - \mu_{0}^{2}
\beta = 0,
\end{equation}
\begin{equation}
e^{2\phi_0} \kappa ( h^{\prime \prime} + \frac{1}{\rho} h^{\prime})
+ (\beta^{\prime})^{2} + \mu_{0}^{2} \beta^2 = 0
\end{equation}
where $\mu_{0} = \lambda e^{2\phi_0}$.
We seek smooth solutions to these equations
for $\rho > 0$ such that $d\tilde{A}$ tends
to zero as $\rho \rightarrow \infty$ and the gravitational field
tends to that of a  cylinder with arbitrary gravitational mass
(which may be zero).
We recall that ${\cal{H}}(u,\rho) = 2\pi \sigma_{0} \, ln\rho$ yields a weak field Newtonian limit corresponding to a cylinder of mass density
$\sigma_0$ per unit length.
 Therefore we require that $h(\rho) \sim C \,ln\rho $ as $\rho \rightarrow \infty$. Such solutions exist for arbitrary $f(u)$ in terms of
modified Bessel functions:
\begin{equation}
\beta(\rho) = K_{0}(\mu_{0}\rho),
\end{equation}
\begin{eqnarray}
\frac{\kappa}{{\mu_{0}}^2} e^{2\phi_0} \,h(\rho) &=& 
\int_{1}^{\rho} \rho^{\prime} ln\rho^{\prime} ( {K_{1}(\mu_{0}\rho^{\prime})}^{2}
+{K_{0}(\mu_{0}\rho^{\prime})}^{2}) d\rho^{\prime} \nonumber \\
& &+ 
\frac{1}{2} \rho^{2} ln\rho \,(K_{0}(\mu_{0}\rho)K_{2}(\mu_{0}\rho)
-(K_{0}(\mu_{0}\rho))^2) + C \,ln\rho.
\end{eqnarray}
where $C$ is an arbitrary non-negative constant.
The profiles $\beta(\rho)$, $h(\rho)-{\mu_0}^{2}\frac{C}{\kappa}
e^{-2\phi_{0}} \,ln\rho$ are displayed in Figure 1 for
$\kappa = \mu_{0} = 1, \phi_{0} = 0$.

The interpretation of these solutions depends on the
form of $f(u)$.
When $f$ is constant the solution is static.
In terms of the coordinates $(t, x, y, z)$ where 
$$ t = \frac{1}{\sqrt{2}}(u+v) \quad , \quad z = \frac{1}{\sqrt{2}} (v-u) \quad, \quad 
x = \rho \cos \psi \quad , \quad
y = \rho \sin \psi$$ we identify $\rho = 0$ as a line source for
$\{{\bf g}, \tilde{A}\}$ along the $z$-axis at each instant.
Writing the field strength $d\tilde{A}$ in terms of 
hyper-electric $e$ and hyper-magnetic $b$ fields  
with respect to $dt$, one finds that a radial $e$ emanates from this
source and it is everywhere transverse to $b$. The fact that
$\int_{C_1} b$ for a closed space-like contour $C_1$ 
and the flux of $e$ through a finite space-like cylinder
depend on the extent of the integration regions
is a reflection of the massive nature of the $\tilde{A}$ field.

When  $f(u)$ is a non-constant bounded function, the solution describes 
a progressive gravitational wave with amplitude $h(\rho)$ 
that propagates together with $\tilde{A}$ 
with amplitude $\beta(\rho)$ in the
$z$ direction at the speed of light. In the other spatial directions 
the $\tilde{A}$ field falls off exponentially to zero at infinity while the
behaviour of the gravitatonal field is determined by $h(\rho)$.
If one interprets the singular domain $\rho = 0$ as a straight wire,
then it acts as a kind of gravitational optical fibre that guides a
pp-type gravitational wave. Such an interpretation has potential astrophysical 
implications.

Given the surprising properties of these exterior self-gravitating
Einstein-Proca solutions
it may be of interest to explore the
generalised Cremmer-Scherk mechanism [7] on the gravitational sector of
low energy effective string theory. It may be of further interest to note that any Einstein-Proca solution can be used to generate a solution
to non-Riemannian theories of gravity [8].
%\newpage
\vskip 5mm

The authors are grateful to TUBITAK for the support of this research.
 RWT is also grateful to the Department of Physics, Middle East Technical University 
for hospitality.
\newpage 
\section{References}

\begin{enumerate}
\item     H. B. Nielsen, P. Olesen, Nucl. Phys. {\bf B61}(1973)45\\
 D. F\"{o}rster, Nucl. Phys. {\bf B81}(1975)84
\item E. Witten, Phys. Lett.{\bf 153B}(1985)243
\\ A. Vilenkin, T. Vachaspati, Phys. Rev. {\bf D35}(1987)1138
\item R. L. Davis, E. P. S. Shellard, Phys. Lett. {\bf 214B}(1988)219
\\ R. L. Davis, E. P. S. Shellard, Phys. Rev. Lett. {\bf 63}(1989)2021
\item B. Carter, Phys. Lett. {\bf 224B}(1989)61
\\ B. Carter, Phys. Lett. {\bf 228B}(1989)466
\\ B. Carter, Nucl. Phys. {\bf B412}(1994)345
\item E. Copeland, M. Hindmarsh, N. Turok, Phys. Rev. Lett. {\bf 58}(1987)1910
\\ E. Copeland, D. Haws, M. Hindmarsh, N. Turok, 
Nucl. Phys. {\bf B306}(1988)908
\item E. Cremmer, J. Scherk, Nucl. Phys. {\bf B72}(1974)117
\\ M. Kalb, P. Ramond, Phys. Rev. {\bf D9}(1974)2273
%\item A. L. Proca, J. de Phys. Rad.{\bf 7}(1936)347 
% \\ S. Deser, Ann. Inst. H. Poincar\'{e},  {\bf AXVI}(1972)79
\item T. Dereli, R. W. Tucker, {\sl String fields and the standard 
model},  hep-th/9808059
\item T. Dereli, M. \"{O}nder, J. Schray, R. W. Tucker, C. Wang,
Class. Q. Grav. {\bf 13}(1996)L103
\end{enumerate}
\end{document}